\documentclass[runningheads]{llncs}
\usepackage{graphicx}
\usepackage{hyperref}
\usepackage{xcolor}

\usepackage{amsmath}
\usepackage{multirow}

\begin{document}
%
\title{Physics-Aware Motion Simulation \\ for 
T2*-Weighted Brain MRI}
\titlerunning{Physics-Aware Motion Simulation for T2*-Weighted Brain MRI}

\author{Hannah Eichhorn\inst{1, 2, *}
\and Kerstin Hammernik\inst{2}
\and Veronika Spieker\inst{1,2}
\and Samira M. Epp\inst{3,4}
\and Daniel Rueckert\inst{2,3,5}
\and Christine Preibisch\inst{3}
\and Julia A. Schnabel\inst{1,2,6}
}

\authorrunning{H. Eichhorn et al.}

\institute{
Institute of Machine Learning in Biomedical Imaging, Helmholtz Munich, Germany,
\and School of Computation, Information and Technology, Technical University of Munich, Germany
\and School of Medicine, Technical University of Munich, Germany
\and Graduate School of Systemic Neurosciences, Ludwig-Maximilians-University, Germany
\and Department of Computing, Imperial College London, United Kingdom 
\and School of Biomedical Engineering and Imaging Sciences, King’s College London, United Kingdom \\
*\email{hannah.eichhorn@helmholtz-munich.de}}

\maketitle              
\begin{abstract}
In this work, we propose a realistic, physics-aware motion simulation procedure for T$_2$*-weighted magnetic resonance imaging (MRI) to improve learning-based motion correction. As T$_2$*-weighted MRI is highly sensitive to motion-related changes in magnetic field inhomogeneities, it is of utmost importance to include physics information in the simulation. Additionally, current motion simulations often only assume simplified motion patterns. Our simulations, on the other hand, include real recorded subject motion and realistic effects of motion-induced magnetic field inhomogeneity changes.
We demonstrate the use of such simulated data by training a convolutional neural network 
to detect the presence of motion in affected k-space lines. 
The network accurately detects motion-affected k-space lines for simulated displacements down to $\geq$0.5~mm (accuracy on test set: $92.5\%$).  
Finally, our results demonstrate exciting opportunities of simulation-based k-space line detection combined with more powerful reconstruction methods.

\keywords{Brain MRI \and Motion Artefacts \and Motion Detection  \and Motion Correction \and Deep Learning.}
\end{abstract}

%
%
%
\section{Introduction}
T$_2$* quantification, as part of the multi-parametric quantitative BOLD (mqBOLD) protocol \cite{Hirsch_2014}, enables oxygenation sensitive brain magnetic resonance imaging (MRI) and is more affordable and less invasive than positron emission tomography techniques. Quantitative MRI, as opposed to the more widely used qualitative structural imaging, allows for a consistent extraction of biomarkers across scanners and hospitals by measuring physical tissue properties.
Promising applications of the quantitative mqBOLD technique comprise stroke, glioma and internal carotid artery stenosis \cite{Gersing_2015,Kaczmarz_2021,Preibisch_2017}. 
Motion artefacts, however, remain a major challenge for brain MRI in general. Specifically, T$_2$*-weighted gradient echo (GRE) MRI, shows a high sensitivity towards magnetic field (B$_0$) inhomogeneities and is hence particularly affected by subject head motion \cite{Magerkurth_2011}. Due to the increasing impact of motion with increasing echo times, motion severely affects the signal decay over several echoes and, thus, the accuracy of the T$_2$* quantification from multi-echo data 
\cite{Noth_2014}. Artefacts have been shown to propagate from the T$_2$* mapping towards derived parameters even for mild motion with less than 1~mm average displacements during the scans \cite{Eichhorn_2023}, which underlines the need of intra-scan motion correction (MoCo) for T$_2$*-weighted GRE data.  

Deep learning solutions have shown promising results for correcting motion in various MR applications. MoCo has been approached as an image denoising problem, using convolutional neural networks \cite{Chatterjee_2020,Xu_2022} as well as generative adversarial networks \cite{Johnson_2019,Kustner_2019}. 
However, acting purely on image data, such methods cannot guarantee consistency with the acquired raw k-space data, which might hinder their translation into clinical practice. 
Enforcing data consistency is only possible when 
combining MoCo with the image reconstruction process \cite{Haskell_2019,Oksuz_2020,Rotman_2021}. 
Yet, the majority of learning-based MoCo techniques - whether purely image-based or combined with the reconstruction process - rely on the availability of paired motion-free and motion-corrupted data for supervised model training \cite{Spieker_2023}. Since the acquisition of large paired datasets is expensive and not always feasible, motion simulation is widely used. However, realistic and physics-aware simulations are still underutilised. Some authors merely exchange k-space lines between different time points \cite{Oksuz_2020}. Others use MR physics as basis for a more realistic simulation, but only combine a small number of discrete motion states, ignoring the continuous nature of real subject motion \cite{Chatterjee_2020,Johnson_2019,Rotman_2021,Xu_2022}; or only simulate in-plane motion instead of realistic 3D motion \cite{Chatterjee_2020,Rotman_2021}. Furthermore, second-order motion effects like motion-induced magnetic field changes are commonly ignored.

In this work, we propose a physics-aware motion simulation procedure that allows for more realistic model training in the presence of magnetic field inhomogeneities and demonstrate its use for the learning-based detection of motion-corrupted k-space lines in T2*-weighted MRI. Our contributions are three-fold: 
\begin{enumerate}
    \item We carry out realistic motion simulations based on real recorded patient motion. In contrast to state-of-the-art methods, we do not only include rigid-body transformations, but also B$_0$ inhomogeneities as second-order effects.
    \item Inspired by the work of Oksuz et al. \cite{Oksuz_2020}, we train a 3D convolutional neural network to classify individual motion-affected k-space lines, leveraging the multi-echo information of the T$_2$*-weighted brain dataset.
    \item To demonstrate the potential of our work, we show motion corrected images where information on motion-affected lines is included as a weighting mask in the data consistency term of the iterative reconstruction procedure.
\end{enumerate}

\section{Methods}

\subsection{Motion Simulation}
The MRI single coil forward model in the presence of rigid-body motion includes the Fourier transform $\mathcal{F}$, the sampling mask $\mathbf{M}_t$ and the three parameter rotation and translation transforms, $\mathbf{R}_t$ and $\mathbf{T}_t$, which are applied to the  motion-free image $x$ for each time point $t$, yielding the motion-affected k-space data $y$ \cite{Atkinson_2023}:
\begin{equation}
\label{eq:MRForward}
    y = \sum_{t=1}^{T} \mathbf{M}_t \mathcal{F} \mathbf{T}_t\mathbf{R}_t x.
\end{equation}
As second-order motion effect, we further include a phase shift, which is induced by position-dependent magnetic field inhomogeneities $\mathbf{\omega_{B_0}}_t$ and increases with the echo time $T_E$:
\begin{equation}
    y = \sum_{t=1}^{T} \mathbf{M}_t \mathcal{F} e^{-2i\pi \mathbf{\omega_{B_0}}_t T_E} \mathbf{T}_t\mathbf{R}_t x.
\end{equation} 

\begin{figure}[t]
    \centering
    \includegraphics[width=115mm]{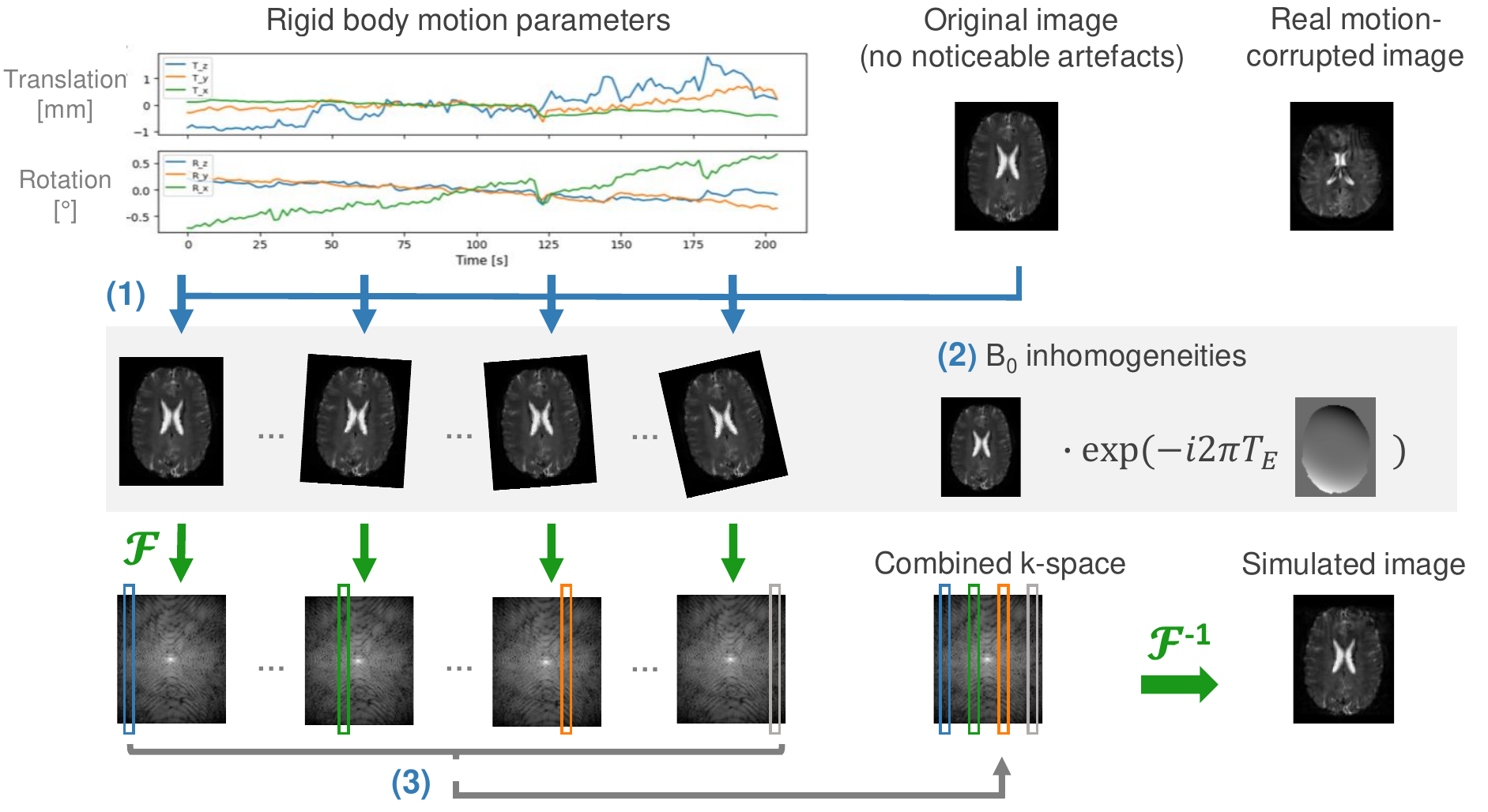}
    \caption{Visualisation of the motion simulation approach. (1) Rigid-body transformations are applied to the complex original image data for each time point. (2) A phase term with randomly generated B$_0$ inhomogeneities is multiplied to the transformed images. The results are transformed into k-space. (3) k-Space lines from the individual motion states 
    are merged into the final simulated k-space, accounting for the multi-slice acquisition scheme. An inverse Fourier transform yields the motion corrupted image. 
    }
    \label{fig:Motion_Sim}
\end{figure}
Based on this, motion is simulated by rigidly transforming the complex MRI data and subsequently merging different motion states in k-space, as visualised in Fig.~\ref{fig:Motion_Sim}. In this process, the actual multi-slice acquisition scheme is considered, i.e. the ordering of phase encoding lines is included in 
$\mathbf{M}_t$. Additionally, for time points with average displacements of more than 0.5~mm, magnetic field inhomogeneities are incorporated by multiplying the image with a B$_0$ map that was modified by adding random image gradients with deviations of max. 5~Hz~\cite{Liu_2018}.

Motion simulation is only performed for phase encoding (PE) lines, where the average displacement of points within a  sphere with radius 64~mm, as model of the head, exceeds a certain threshold value $d_{min}$. To investigate the amount of motion that can be detected, this so-called \textit{simulation threshold} is varied between 0.25~mm and 1.0~mm in the experiments.
To avoid the need of registration for calculating full-reference metrics with respect to the original motion-free image,
the motion curves are transformed in a way that the median motion state - measured by the average displacement - is in the zero-position.

\subsection{k-Space Line Classification Network}
The line classification network is adapted from Oksuz et al. \cite{Oksuz_2020} and consists of five repetitions of 3D convolution, batch normalisation, ReLU activation, dropout and max pooling layers, followed by a fully connected, a dropout and a sigmoid activation layer, as visualised in Fig.~S1 of the Supplementary Material. Convolution layers are implemented with a kernel size of $3\times3\times3$. Max pooling is performed in the echo and readout dimensions. 
The network is trained for 300 epochs with Adam~\cite{Adam_2015} using a learning rate of $5 \times 10^{-4}$ and a weighted cross entropy loss between prediction $m^{pr}$ and target $m^{ta}$:
\begin{equation}
    L(m^{pr}, m^{ta}) = - m^{ta} \log (m^{pr} + \epsilon) \ - \ \omega \cdot (1-m^{ta}) \log((1-m^{pr}) + \epsilon).
\end{equation} 
Weights of the epoch with the lowest validation loss are used.
Batch sizes of 32 and 64, weighting factors, $\omega$, between 1 and 5, weight decays between $1 \times 10^{-5}$ and $5 \times 10^{-2}$ and dropout probabilities between 10$\%$ and 30$\%$ are tested and the best configuration is chosen based on the validation dataset (64 / 5 / $1 \times 10^{-3}$ /  20$\%$; 219,380 trainable parameters). The average runtime of the network training is 14 hours.
Computations are performed on an NVIDIA RTX A6000, using  Python 3.8.12 and PyTorch 1.13.0 (code available at: \url{https://github.com/HannahEichhorn/T2starLineDet}).

\subsection{Data}
Motion simulation for training and evaluation of the line classification network is performed on 116 complex, coil-combined T$_2$*-weighted datasets of 59 healthy volunteers, which can be expected to not considerably move during the scanning. We visually confirmed that these data did not show noticeable motion artefacts. The unpublished mqBOLD data originates from four ongoing studies investigating brain oxygen metabolism on a 3T Philips Elition MR scanner (Philips Healthcare, Best, The Netherlands), using a multi-slice GRE sequence (12 echoes, 30-36 slices, TE1/$\Delta$TE = 5/5~ms, TR=1910-2291~ms, $\alpha$=30°, voxel size: $2\times2\times3$~mm$^3$). The ongoing studies have been approved by the local ethics committee (approval numbers 472/16 S, 446/21 S, 165/21 S, 382/18 S).
The available datasets are divided subject-wise into train, validation and test sets (74/18/24 datasets). 

For a realistic motion simulation, we base our simulation on real head motion. We extract 132 motion curves of length 235~s from 62 functional MRI (fMRI) time series (unpublished data from two ongoing studies, independent cohorts from above imaging cohorts). 
These are divided into train, validation and test sets ($N=$ 88/20/24 motion curves), keeping the amount of motion as equal as possible in the different sets. Principal component analysis is used to determine the largest modes of variations of the training motion curves $pc_i(t)$, which are combined with random weights $\alpha_i$ and added to the mean curve $\overline{T(t)}$ to generate augmented training samples~\cite{Cootes_1995}:
\begin{equation}
    T_{new}(t) = \overline{T(t)} + \sum_{i=1}^{0.2\cdot N} \alpha_i \cdot pc_i(t).
\end{equation}
For each complex image in the training set, six motion curves are generated for simulation. Slices with 
more than 30$\%$ background voxels are excluded. This results in a total of 8,922 slices for training, 392 for validation and 492 for testing.

\noindent \textbf{Preprocessing:} The complex data are normalised for each line individually: 
\begin{equation}
    y^{n}_{epr} =  \frac{y_{epr}}{\sqrt{\sum_{ep} \mid y_{epr} \mid ^2}},
\end{equation}
with $e$, $p$ and $r$ indicating the echo, phase encoding and readout dimension. Real and imaginary parts are fed into the network as different channels. 

The target classification labels are calculated by thresholding and inverting the average displacement of a sphere with a 64~mm radius at the acquisition time of the corresponding PE line, using the simulation threshold value $d_{min}$. The target masks are averaged across all 12 echoes, since these have been acquired within 60~ms and are thus assumed to have been acquired during the same motion state.

\subsection{Evaluation}
The predicted classification labels are evaluated based on accuracy, corresponding to the rate of correctly predicted lines.
For a more detailed interpretation, the rate of non-detected lines (ND) is calculated as the fraction of lines with motion ($m^{ta}=0$), which the network does not detect ($m^{pr}=1$). Similarly, the rate of wrongly-detected lines (WD) is calculated as the fraction of lines without motion ($m^{ta}=1$), which the network classifies as motion-corrupted ($m^{pr}=0$).

\subsection{Weighted Iterative Reconstruction}
Inspired by the work of Jiang et al. \cite{Jiang_2018}, we include the estimated motion classification as weights into the data consistency (DC) term of the single coil MRI reconstruction model with total variation (TV) regularisation:
\begin{equation}
    x^* = \arg \min_{x} \frac{1}{2} \parallel \mathbf{W} (\mathbf{A}x-y) \parallel_2^2 + \lambda \parallel \mathbf{\Phi} x \parallel_1.
    \label{eq:TVrecon}
\end{equation}
$\mathbf{W}$ represents a diagonal matrix with empirical weights, 0.25 or~1 for the presence or absence of motion in each PE line, 
$\mathbf{A} = \mathcal{F}$ the forward operator, $\lambda=2$ the regularisation parameter obtained empirically and $\mathbf{\Phi}$ the finite differences operator. Eq.~\ref{eq:TVrecon} can be solved for the optimal $x^*$ with a proximal gradient algorithm~\cite{Hammernik_2020}.

\section{Experiments and Results}
\begin{figure}[t]
    \centerline{\includegraphics[width=115mm]{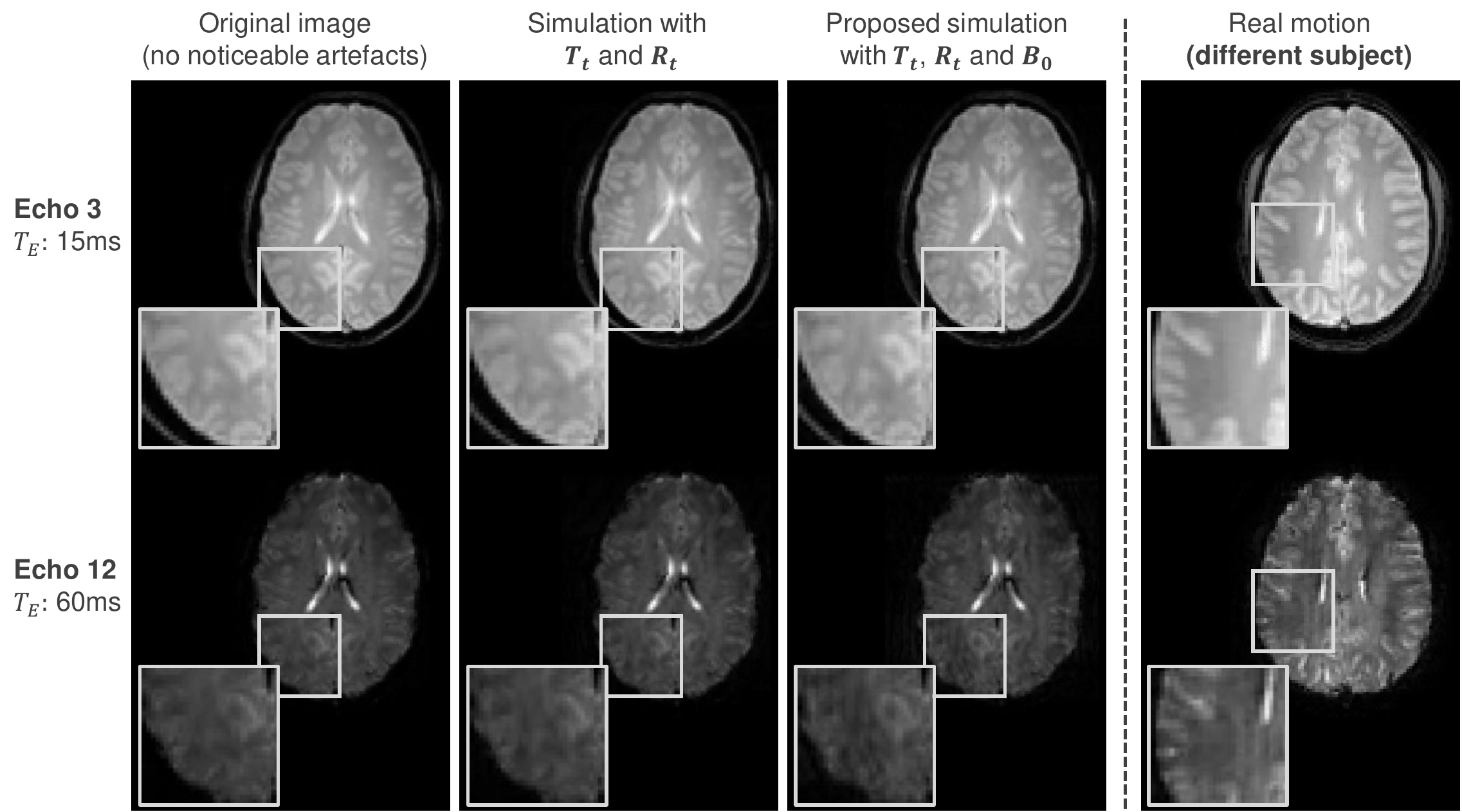}}
    \caption{Example images showing the relevance of B$_0$ inhomogeneities for generating realistic artefact patterns. From left to right:  original images (without noticeable motion artefacts), simulated images without and with B$_0$ inhomogeneities and real motion images for two different echo times (top and bottom row). The motion curve for the simulation is shown in Fig.~S2 of the Supplementary Material (mean displacement during the whole scan: 0.89~mm, simulation threshold: 0.5~mm). Artefacts - i.e. blurring of underlying structures and wave-like patterns - are more pronounced for later echoes in the real motion case and in the proposed simulation method.
    }
    \label{fig:Example_simulations}
\end{figure}

In Fig.~\ref{fig:Example_simulations} we visually compare simulations of motion-affected images without and with B$_0$ inhomogeneities with a real motion case. This example demonstrates the relevance of including motion-induced field inhomogeneity changes for generating realistic artefact patterns, i.e. more severe blurring and wave-like artefacts appearing for later echoes.

\begin{figure}[t]
    \centerline{\includegraphics[width=120mm]{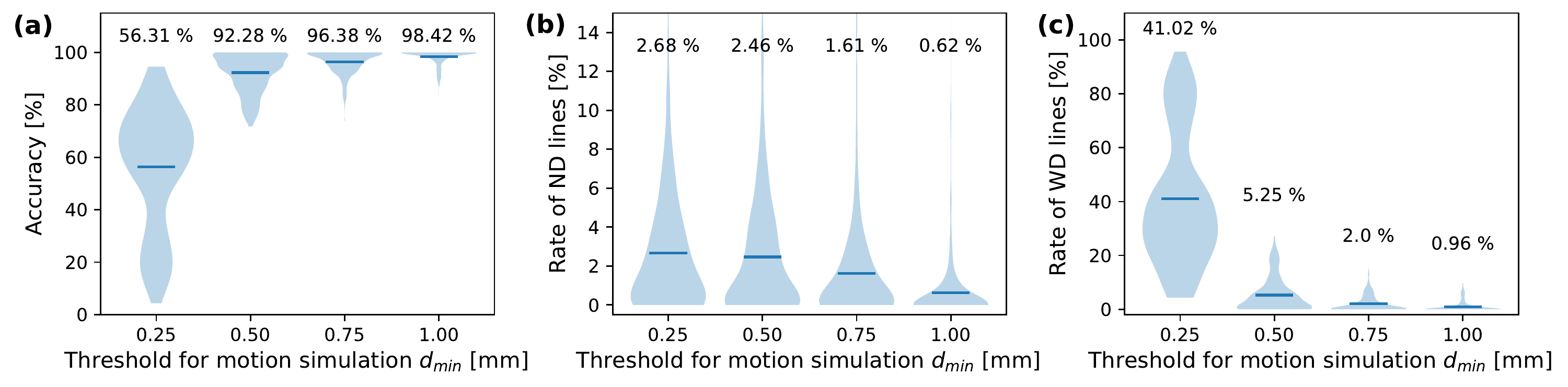}}
    \caption{(a) Test accuracy, (b) rates of non-detected (ND) and (c) wrongly-detected (WD) k-space lines for varying thresholds of simulated motion in train and test data. Mean values are visualised by horizontal lines and given in numbers. All metrics show a decreasing classification performance with decreasing simulation thresholds.}
    \label{fig:Performance_different_thresholds}
\end{figure}

In the following, we use the simulated data to investigate the minimal level of simulated motion, which can be detected by the line classification network. We train the network using four different datasets which were simulated with varying simulation thresholds $d_{min}$. Fig.~\ref{fig:Performance_different_thresholds} compares accuracy, ND and WD rates of these networks' predictions on unseen test data (with the same simulation thresholds as for training). 
The performance in terms of accuracy, ND and WD rate decreases for training and testing the networks with decreasing simulation thresholds. When changing the simulation threshold from 0.5~mm to 0.25~mm, the accuracy drops from $92.28\%$ to $56.31\%$, which is mainly driven by an increase of the WD rate from $5.25\%$ to $41.02\%$. This indicates that the network is not able to correctly classify motion-affected PE lines when motion is simulated for time points with less than 0.5~mm average displacement of a 64~mm sphere. 

\begin{figure}[t]
    \centering
    \includegraphics[width=115mm]{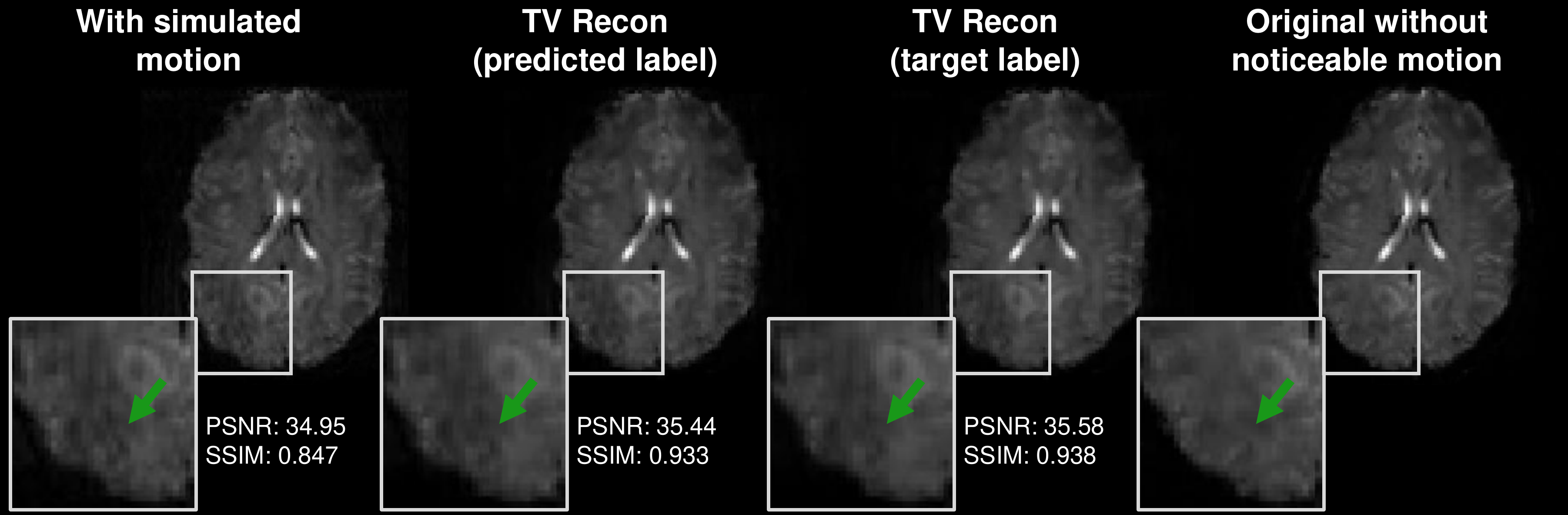}
    \caption{Demonstration of weighted reconstructions 
    for simulated data. 
    From left to right: images with simulated motion  (same data as in Fig.~\ref{fig:Example_simulations}, simulation threshold: 0.5~mm), TV reconstructions weighted with predicted labels by the network as well as target labels, and original images. 
    Peak signal-to-noise ratio (PSNR) and structural similarity index (SSIM) \cite{Wang_2004} with respect to the original images are given below the images. Green arrows indicate an area with subtly reduced artefacts in the weighted reconstructions.}
    \label{fig:Example_Recons}
\end{figure}

As an outlook, Fig.~\ref{fig:Example_Recons} demonstrates that the predicted line-wise classification labels can be used for correcting motion-corrupted images by including the classifications as weights in the DC term of the reconstruction model 
(Eq.~\ref{eq:TVrecon}).
 Visual inspection reveals a subtle decrease of motion artefacts in the weighted reconstruction with target labels and to a certain extent also in the reconstruction with predicted labels. This is in agreement with an improvement of quantitative metrics for both scenarios. However, the weighted reconstructions appear slightly over-smooth compared to the original image.

\section{Discussion}
We have proposed a physics-aware method to realistically simulate motion artefacts for T$_2$*-weighted brain MRI in the presence of magnetic field inhomogeneities, which enables supervised training of MoCo models, and, thus, avoids the expensive acquisition of paired in-vivo data. We demonstrated the use of such simulated data for training a k-space line detection network and showed a usecase of this approach for a motion-corrected reconstruction.

Most learning-based MoCo approaches rely on simplistic motion simulation for supervised model training. Oksuz et al. \cite{Oksuz_2020}, for instance, simulate motion in dynamic cardiac MRI as mistriggering artefacts by exchanging k-space lines between different time points, which does not cover the full range of possible motion artefacts. Others use the MR forward model (Eq.~\ref{eq:MRForward}) as basis for a more realistic simulation, but only merge a few, discrete motion states and, thus, ignore the continuous nature of real subject motion \cite{Chatterjee_2020,Johnson_2019,Rotman_2021,Xu_2022}; 
or only simulate in-plane motion instead of realistic 3D motion \cite{Chatterjee_2020,Rotman_2021}.
To improve the performance of the developed algorithms on real motion data, the training data generation needs to be as realistic as possible. Thus, our proposed simulation is based on real recorded 3D subject head motion extracted from fMRI time series. Furthermore, due to the sensitivity of T2*-weighted GRE to magnetic field inhomogeneities, our physics-aware simulation incorporates B$_0$ inhomogeneity changes as secondary motion effects in addition to rigid-body transformations.

The comparison of simulation examples with a real motion case in Fig.~\ref{fig:Example_simulations} demonstrated that the proposed inclusion of B$_0$ changes in the simulation results in more plausible artefact patterns. Note that - due to inherently long acquisition times - the resolution of quantitative MRI (with typical voxel sizes of 2-3~mm) is commonly lower than for qualitative structural MRI (e.g. T1-weighted MPRAGE scans with voxel sizes $<$ 1~mm).  
Furthermore, the shown motion artefacts, which mainly manifest in blurring and wave-like patterns, might not appear as severe as in previous studies, where authors simulated motion considerably larger than a voxel size \cite{Haskell_2019,Johnson_2019,Kustner_2019}.
However, MoCo is required even for more cooperative subjects in the context of T$_2$* quantification based on multi-echo GRE data. 
That is because motion and associated B$_0$ changes disturb the signal decay over several echo times, even for displacements smaller than the voxel size of 2~mm~\cite{Eichhorn_2023}.  

To showcase the use of the proposed simulation, we trained a convolutional neural network to detect the presence or absence of motion for every k-space line. First, we have investigated the extent of motion that can be detected by the network. For this, we have varied the simulation threshold for both training and test data. The network was able to classify displacements down to 0.5~mm, corresponding to one quarter of the voxel size, with an accuracy of $92.3\%$. For smaller simulation thresholds, we have observed a clear performance drop.  
Considering that the underlying motion curves were extracted from real fMRI time series, a certain degree of noise due to the co-registration can be expected for very small displacements.  
Thus, using such motion information in the simulation might lead to an overestimation of motion artefacts, which justifies to simulate motion only for time points with displacements larger than 0.5~mm.

As proof-of-principle, we have included the classification labels as weights in the DC term of an iterative reconstruction and demonstrated how the detected motion information can be used for MoCo. 
The results showed subtly reduced motion artefacts, both visually and quantitatively. However, the reconstructions appeared slightly over-smoothed compared to the original image, which is a common problem when using TV regularisation. 
In future work we plan to combine line-wise classifications as DC weights with e.g. learning an unrolled optimisation scheme~\cite{
Schlemper_2018}. We further plan to utilise multi-coil raw k-space data to exploit redundancies between coils for improved reconstruction performance. 

One limitation of the proposed motion simulation procedure is that we generated random B$_0$ inhomogeneity variations for each position. More authentic, pose-dependent B$_0$ inhomogeneity maps can be synthesized by interpolating between B$_0$ fields of the same subject acquired in different positions \cite{Brackenier_2022}. This requires repeated measurements of each subject in different positions, though, which is not feasible for generating large-scale training datasets.
Finally, an evaluation of the trained network using real motion corrupted data remains for future work. We just received ethics approval to perform paired in-vivo experiments, which will allow for more thorough quantitative evaluations.

\subsubsection*{Conclusion}
We have presented a method for realistic, physics-based motion simulation in T2*-weighted GRE MRI. Our proposed method uses motion curves extracted from real, time-resolved fMRI data and includes motion induced changes in B$_0$ inhomogeneities. We demonstrated the utility of these simulations for training a line-wise classification network and have been able to detect displacements down to 0.5~mm (a quarter of a voxel). Finally, we outlined how the simulation-based motion detection might be included in the data consistency term of an iterative reconstruction procedure, providing a promising research direction for future work.

\section*{Acknowledgements}
V.S. and H.E. are partially supported by the Helmholtz Association under the joint research school ”Munich School for Data Science - MUDS”.

%
%
%
\bibliographystyle{splncs04}
\bibliography{references}

\section*{Supplementary Material}
\setcounter{figure}{0}
\renewcommand{\figurename}{Fig.}
\renewcommand{\thefigure}{S\arabic{figure}}

\begin{figure}[h!]
    \centering
    \includegraphics[width=110mm]{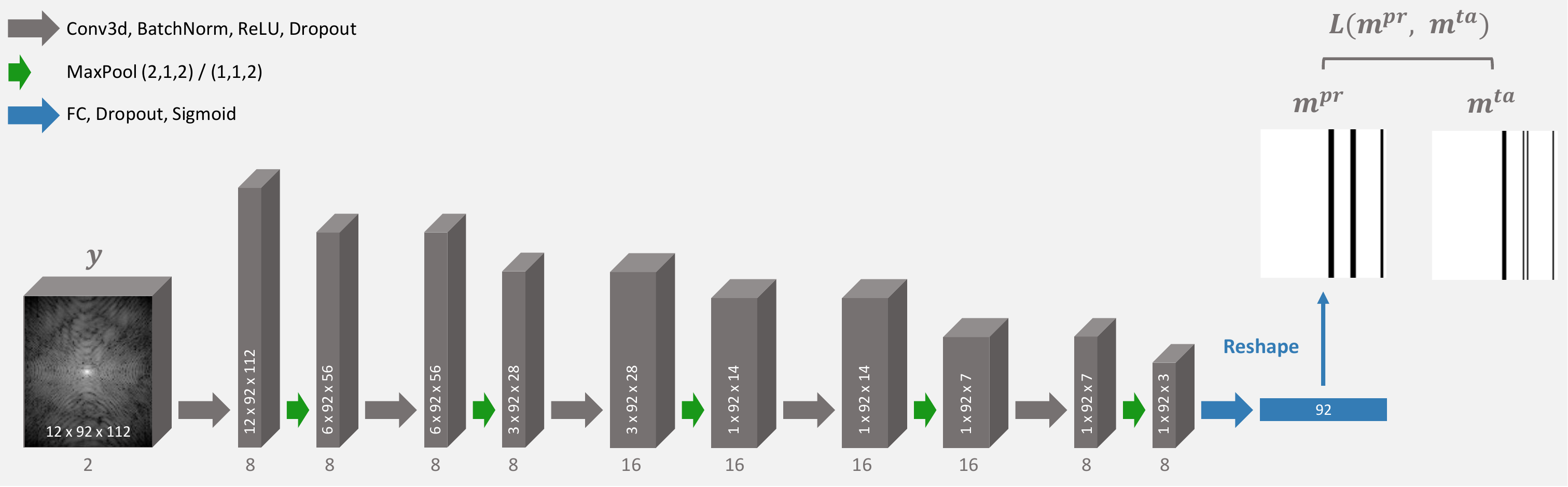}
    \caption{Network architecture for line-wise classification. Input $y$ of the network is the motion corrupted k-space. The 219,380 trainable parameters are updated based on the loss calculated between the predicted and the target classification, $m^{pr}$ and $m^{ta}$, which are visualised in black or white for the presence or absence of motion in each PE line.}
    \label{fig:Network_architecture}
\end{figure}

\begin{figure}[h!]
    \centerline{\includegraphics[width=120mm]{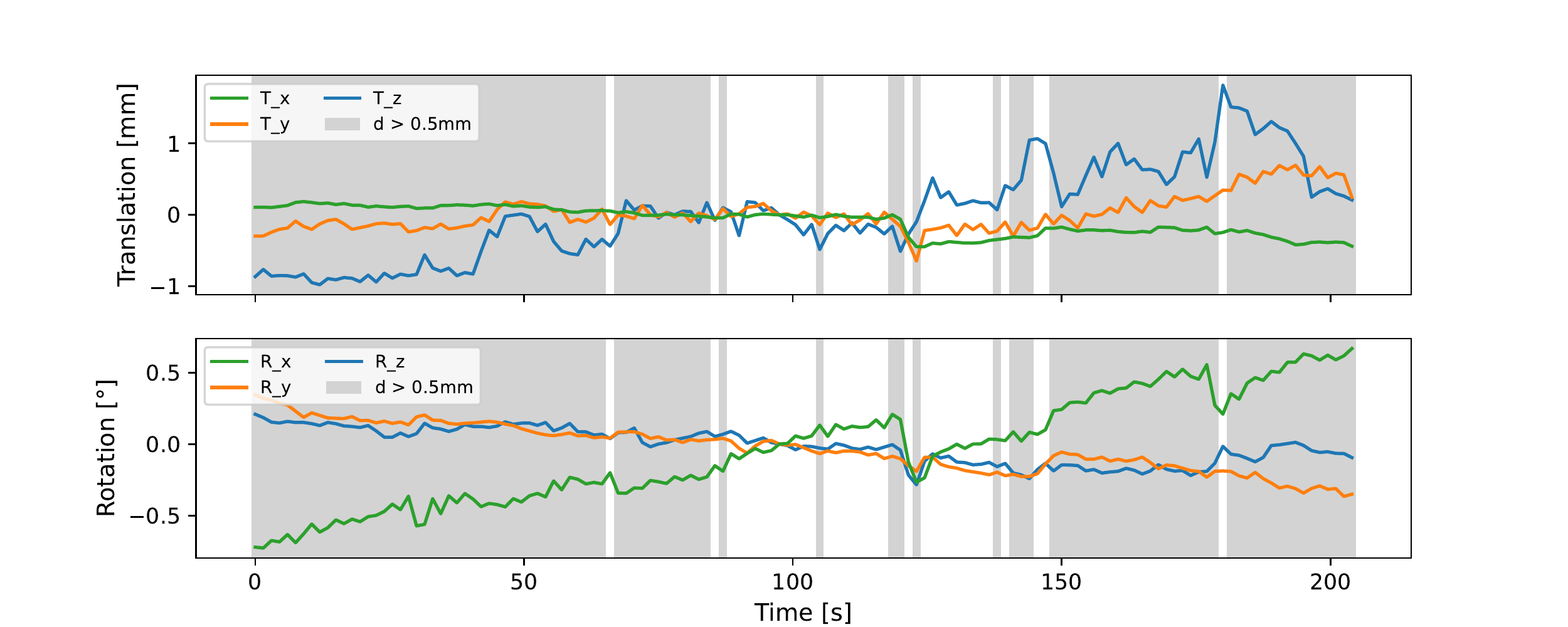}}
    \caption{Motion parameter curves used in the simulation of the example images shown in Figs.~2 and 4 of the main article (mean displacements during the whole scan: 0.89~mm).
    Translations along x-, y- and z-axis are shown in the top row, rotations around x-, y- and z-axis in the bottom row. Grey background represents time points with displacements, d, larger than 0.5~mm, which corresponds to time points, for which motion was simulated.}
    \label{fig:Suppl_Motion_curves}
\end{figure}

\end{document}